\documentclass[12pt, a4paper, twoside]{scrartcl}
\pdfoutput=1
\usepackage{lineno}
\usepackage{todonotes}

\usepackage[tracking=true,expansion=true,stretch=15,shrink=15]{microtype}
\DeclareMicrotypeSet*[tracking]{my}
{ font = */*/*/sc/* } 
\SetTracking{ encoding = *, shape = sc }{ 45 }
\SetProtrusion{encoding={*},family={bch},series={*},size={6,7}}
{1={ ,750},2={ ,500},3={ ,500},4={ ,500},5={ ,500},
	6={ ,500},7={ ,600},8={ ,500},9={ ,500},0={ ,500}}
\SetExtraKerning[unit=space]
{encoding={*}, family={qhv}, series={b}, size={large,Large}}
{1={-200,-200}, 
	\textendash={400,400}}

\usepackage{setspace}

\usepackage{latexsym}
\usepackage{rotating}
\usepackage{amssymb}
\usepackage{amsmath}
\usepackage{amsthm}
\usepackage{shadethm}
\usepackage{dsfont}
\usepackage{mathrsfs}
\usepackage{empheq}
\usepackage{bm}
\usepackage{subfig}
\usepackage{enumerate}
\usepackage{graphicx}
\usepackage{algorithm}
\usepackage{algpseudocode}
\usepackage{pdfpages}
\usepackage[round,comma]{natbib}
\usepackage[automark,headsepline]{scrpage2}
\usepackage{hyperref}
\usepackage{nameref}
\definecolor{TUMblue}{cmyk}{1, .54, .04, .19}
\hypersetup{
	colorlinks   = true, %Colours links instead of ugly boxes
	urlcolor     = TUMblue, %Colour for external hyperlinks
	linkcolor    = TUMblue, %Colour of internal links
	citecolor   = TUMblue %Colour of citations
}

\usepackage[
left=1.3in,
right=1.3in,
top=0.7in,
bottom=0.7in,
includeheadfoot
]{geometry}

\usepackage{tikz}
\usetikzlibrary{shapes,arrows,decorations.pathmorphing,graphs,positioning,backgrounds}

\newtheoremstyle{new}{12pt}{12pt}{\itshape}{}{\bfseries}{.}{1em}{}
\theoremstyle{new}

\newtheorem{Example}{Example}

\newcommand{\R}{\mathds{R}}

\newcommand{\wh}{\widehat}

\newcommand{\bv}{{\boldsymbol v}}

\newcommand{\lshort}{\lambda}

\newcommand*\patchAmsMathEnvironmentForLineno[1]{%
	\expandafter\let\csname old#1\expandafter\endcsname\csname #1\endcsname
	\expandafter\let\csname oldend#1\expandafter\endcsname\csname end#1\endcsname
	\renewenvironment{#1}%
	{\linenomath\csname old#1\endcsname}%
	{\csname oldend#1\endcsname\endlinenomath}}% 
\newcommand*\patchBothAmsMathEnvironmentsForLineno[1]{%
	\patchAmsMathEnvironmentForLineno{#1}%
	\patchAmsMathEnvironmentForLineno{#1*}}%
\AtBeginDocument{%
	\patchBothAmsMathEnvironmentsForLineno{equation}%
	\patchBothAmsMathEnvironmentsForLineno{align}%
	\patchBothAmsMathEnvironmentsForLineno{flalign}%
	\patchBothAmsMathEnvironmentsForLineno{alignat}%
	\patchBothAmsMathEnvironmentsForLineno{gather}%
	\patchBothAmsMathEnvironmentsForLineno{multline}%
}

\begin{document}

	\pagestyle{scrheadings}
	\clearscrheadings
	\lohead{Nonparametric estimation of simplified vines: comparison of methods}
	\rohead{\pagemark}
    \lehead{T.\ Nagler, C.\ Schellhase, C.\ Czado}
    \rehead{\pagemark}

    \raggedbottom
    
	% content
    \author{Thomas Nagler\thanks{Corresponding author, Department of Mathematics, Technische Universit\"at M\"unchen, Boltzmanstra{\ss}e 3, 85748 Garching (email: thomas.nagler@tum.de)}, Christian Schellhase\thanks{Centre for Statistics, Bielefeld University, Department of Business Administration and Economics, Germany. email: cschellhase@wiwi.uni-bielefeld.de}, Claudia Czado\thanks{Department of Mathematics, Technische Universit\"at M\"unchen, Boltzmanstra{\ss}e 3, 85748 Garching (email: cczado@ma.tum.de)}}

\title{Nonparametric estimation of simplified vine copula models:
comparison of methods}
\date{\hspace{3pt} \normalsize\today}

\maketitle

\begin{abstract} 
\noindent {\bfseries \sffamily Abstract}\\

\noindent In the last decade, simplified vine copula models have been an active area of research. They build a high dimensional probability density from the product of marginals densities and bivariate copula densities. Besides parametric models, several approaches to nonparametric estimation of vine copulas have been proposed. In this article, we extend these approaches and compare them in an extensive simulation study and a real data application. We identify several factors driving the relative performance of the estimators. The most important one is the strength of dependence. No method was found to be uniformly better than all others. Overall, the kernel estimators performed best, but do worse than penalized B-spline estimators when there is weak dependence and no tail dependence.  \\[12pt]
	{\itshape Keywords:  B-spline, Bernstein, copula, kernel, nonparametric, simulation, vine}
\end{abstract}
	\section{Introduction} 
\label{sec:introduction}
Simplified vine copulas, or pair-copula constructions (PCC), have become very popular over the last decade \citep{Bedford02,Aas09,Czado10,Kurowicka10,Brechmann12,Spanhel2015}. Vine copula models build a high-dimensional dependence structure by hierarchical modeling of bivariate copulas (the \emph{pair-copulas}). Each pair-copula can be specified as a unique parametric copula function. Thus, simplified vine copula models give rise to very flexible models which are often found to be superior to other multivariate copula models \citep{Aas09,Fischer2009}. The models are also easily tractable because pair-copulas can be estimated sequentially. Parametric models for the pair-copulas are most common, but bear the risk of wrong specifications. In particular, most parametric families only allow for highly symmetric and monotone relationships between variables. 

To remedy this issue, several nonparametric approaches have been proposed: \citet{ScheKau:12b} use penalized Bernstein polynomials and B-splines,  \citet{Nagler15} use kernel estimators, and \citet{SchefferWeiß16} use a non-penalized Bernstein estimator. A related contribution of \citet{Haff15} introduces the empirical pair-copula as an extension of the empirical copula, but does not aim at estimation of the vine copula density which is the focus of this article.

From a practitioner's point of view, the question arises: which method should I choose for a given data set? This question is difficult to answer theoretically because asymptotic approximations of nonparametric vine copula density estimators are prohibitively unwieldy \citep[cf.,][Propositions 2 and 5]{Nagler15}. In this article, we conduct an extensive simulation study to provide some guidance nevertheless. All estimation methods will be compared under several specifications of strength and type of dependence, sample size, and dimension, thereby covering a large range of practical scenarios. 

Although our primary goal is to survey and compare existing methods, we extend the estimators proposed by \citet{ScheKau:12b}, \citet{SchefferWeiß16}, and \citet{Nagler15} in several ways:
\begin{itemize}
\item The estimators of \citet{ScheKau:12b} and \citet{SchefferWeiß16} are extended to allow for general R-vine structures (opposed to just D- and/or C-vine structures).
\item Besides linear B-splines \citep[as in][]{ScheKau:12b}, we also consider quadratic B-splines.
\item Beyond the classical kernel density estimator used in \citet{Nagler15}, we further consider local linear and local quadratic likelihood kernel estimators.
\item All pair-copula estimators can be combined with structure selection algorithms using both Kendall's $\tau$ and a corrected AIC as target criterion.
\end{itemize}

The remainder of this article is organized as follows. \autoref{sec:vines} introduces simplified vine copula models. \autoref{sec:inference} presents and extends several existing nonparametric methods for pair-copula estimation, describes a step-wise estimation algorithm  for vine-copula estimation, and discusses approaches for model selection. We describe the  design of our simulation study in \autoref{sec:simulations} and summarize the results in \autoref{sec:results}. In \autoref{sec:illustration}, a real data set is used to illustrate the estimators' behavior and demonstrate the necessity for nonparametric estimators. \autoref{sec:conclusion} contains our conclusions.
    \section{Background on simplified vine copula models}
\label{sec:vines}

This section gives a brief introduction to pair-copula constructions. For a more extensive treatment, we refer to \citet{Aas09}, \cite{Czado10}, and \citet{Joe14}.

By Sklar's theorem \citep{Sklar59}, any multivariate distribution function $F$ can be split into its marginal distributions $F_1, \dots, F_d$ and a \emph{copula} $C$:
\begin{align*}
    F(x_1, \dots, x_d) = C\bigl(F_1(x_1), \dots, F_d(x_d)\bigr)
\end{align*}
The copula $C$ describes the dependence structure of the random vector $\bm X$. It is, in fact, the joint distribution of the random vector $\bm U = (U_1, \dots, U_d) = \bigl(F_1(X_1), \dots, F_d(X_d)\bigr)$. Note that $U_1, \dots, U_d$ are uniformly distributed on the unit interval. If $F$ admits a density, we can differentiate the above equation to get
\begin{align}
    f(x_1, \dots, x_d) = c\bigl(F_1(x_1), \dots, F_d(x_d)\bigr) \times \prod_{k=1}^d f_k(x_k),
    \label{sec:vines:sklar_density_eq}
\end{align}
where $c, f_1, \dots, f_d$ are the probability density functions corresponding to $C$,\linebreak$ F_1, \dots, F_d$ respectively. 

Following  \citet{Joe97} and \citet{Bedford01, Bedford02}, any copula density $c$ can be decomposed into a product of $d(d-1)/2$ bivariate (conditional) copula densities. The decomposition is not unique, but all possible decomposition can be organized as graphical structure, called \emph{regular vine (R-vine)}. It is a sequence of trees $\mathcal{V} = T_m = (V_m, E_m), m = 1, \dots, d-1$ satisfying the following conditions:
	\begin{enumerate}
		\item $T_1$ is a tree with nodes $V_1=\{1, \dots, d\}$ and edges $E_1$.
		\item For $m\ge 2$, $T_m$ is a tree with nodes $V_m=E_{m-1}$ and edges $E_m$.
		\item (\emph{Proximity condition}) Whenever two nodes in $T_{m+1}$ are joined by an edge, the corresponding edges in $T_m$ must share a common node.
	\end{enumerate}   

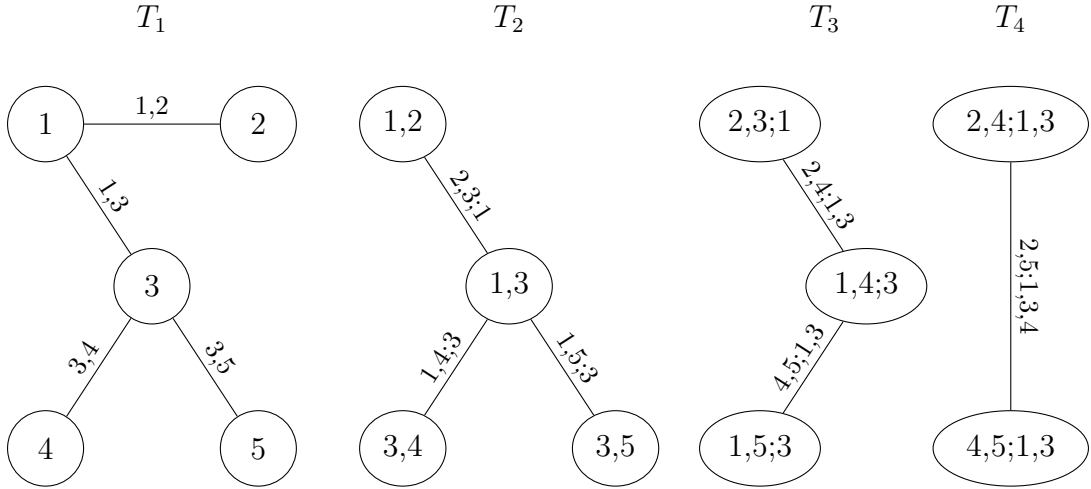
\begin{figure}
\tikzstyle{VineNode} = [ellipse, fill = white, draw = black, text = black, align = center, minimum height = 1cm, minimum width = 1cm]
\tikzstyle{DummyNode}  = [draw = none, fill = none, text = white] % white node without text for better positioning
\tikzstyle{TreeLabels} = [draw = none, fill = none, text = black] % T_1, T_2, etc.
\newcommand{\labelsize}{\footnotesize} % Size of labels
\newcommand{\yshiftLabel}{-0.3cm}  % Reduce space between labels and eges
\newcommand{\yshiftNodes}{-0.75cm} % Reduce space between lines
\newcommand{\xshiftTree}{0.5cm}    % Space between trees
\newcommand{\rotateLabels}{-57}    % Rotation of labels
	\centering
	\begin{tikzpicture}	[every node/.style = VineNode, node distance =1.4cm] % node distance = scaling of edges
	%%% Tree 1 nodes
	\node (1){1}
	node[DummyNode]  (Dummy12)   [right of = 1]{} 
	node             (2)         [right of = Dummy12] {2}
	node             (3)         [below of = Dummy12, yshift = \yshiftNodes] {3}
	node[DummyNode]  (Dummy45)   [below of = 3, yshift = \yshiftNodes]{}
	node             (4)         [left of = Dummy45] {4}
    node             (5)         [right of = Dummy45] {5}
    %%% Tree 2 nodes
    node             (12)        [right of = 2, xshift = \xshiftTree] {1,2} % 
    node[DummyNode]  (Dummy12x)  [right of = 12]{} 
	node             (13)        [below of = Dummy12x, yshift = \yshiftNodes] {1,3}
	node[DummyNode]  (Dummy45c3) [below of = 13, yshift = \yshiftNodes]{} % Notation c fuer conditioned
	node             (34)        [left  of = Dummy45c3] {3,4}
    node             (35)        [right of = Dummy45c3] {3,5}
    %%% Tree 3 nodes
    node             (15c3)      [right of = 35, xshift = \xshiftTree] {1,5;3}
	node[DummyNode]  (Dummy15c3x)[right of = 15c3]{}
	node             (14c3)      [above of = Dummy15c3x, yshift = -\yshiftNodes] {1,4;3}
	node[DummyNode]  (Dummy23c1x)[above of = 14c3, yshift = -\yshiftNodes]{}
	node             (23c1)      [left of  = Dummy23c1x] {2,3;1}	
	%%% Tree 4 nodes
	node             (24c13)     [right of = Dummy23c1x, xshift = \xshiftTree] {2,4;1,3}
    node             (45c13)     [right of = Dummy15c3x, xshift = \xshiftTree] {4,5;1,3}
    %%% Tree labels
	node[TreeLabels] (T1)        [above of = Dummy12] {$T_1$}
	node[TreeLabels] (T2)        [above of = Dummy12x] {$T_2$}
	node[TreeLabels] (T3)        [above of = 23c1] {\hspace{1.7cm}$T_3$} % T_3 manuell mittiger ausrichten (ausprobieren)		
	node[TreeLabels] (T4)        [above of = 24c13] {$T_4$}	 
	;	    	
	%%% Tree 1 edges   
	\draw (1) to node[draw=none, fill = none, font = \labelsize,
	                    above, yshift = \yshiftLabel] {1,2} (2);
	\draw (1) to node[draw=none, fill = none, font = \labelsize, 
	                    rotate = \rotateLabels, above, yshift = \yshiftLabel] {1,3} (3);   
	\draw (3) to node[draw=none, fill = none, font = \labelsize, 
	                    rotate = \rotateLabels, above, yshift = \yshiftLabel] {3,5} (5);  
	\draw (3) to node[draw=none, fill = none, font = \labelsize, 
	                    rotate = -\rotateLabels, above, yshift = \yshiftLabel] {3,4} (4); 
	%%% Tree 2 edges  
	\draw (12) to node[draw=none, fill = none, font = \labelsize, above, 
	                    rotate = \rotateLabels, above, yshift = \yshiftLabel] {2,3;1} (13);   
	\draw (13) to node[draw=none, fill = none, font = \labelsize, above, 
	                    rotate = \rotateLabels, above, yshift = \yshiftLabel] {1,5;3} (35); 
	\draw (13) to node[draw=none, fill = none, font = \labelsize, above, 
	                    rotate = -\rotateLabels, above, yshift = \yshiftLabel] {1,4;3} (34); 
	%%% Tree 3 edges  
	\draw (23c1) to node[draw=none, fill = none, font = \labelsize, above, 
	                    rotate = \rotateLabels, above, yshift = \yshiftLabel] {2,4;1,3} (14c3);   
	\draw (14c3) to node[draw=none, fill = none, font = \labelsize, above, 
	                    rotate = -\rotateLabels, above, yshift = \yshiftLabel] {4,5;1,3} (15c3); 
	%%% Tree 4 edges 
	\draw (24c13) to node[draw=none, fill = none, font = \labelsize, above, 
	                    rotate = -90, above, yshift = \yshiftLabel] {2,5;1,3,4} (45c13);
	\end{tikzpicture}
	\caption{Example of a regular vine tree sequence.}
	\label{fig:trees}
\end{figure}

\autoref{fig:trees} shows an example of a regular vine with each edge $e$ annotated by $(j_e,k_e;D_e)$. The notation for an edge $e$ in $T_i$ depends on the two shared edges in $T_{i-1}$, denoted by $a=(j_a,k_a;D_a)$ and $b=(j_b,k_b;D_b)$ with ${\mathcal{V}_a}=\{j_a,k_a,D_a\}$ and ${\mathcal{V}_b}=\{j_b,k_b,D_b\}$. Here $D_e$ is   called conditioning set while $\{j_e,k_e\}$ is the conditioned set of an edge $e$. In Tree $T_i$, the nodes $a$ and $b$ are joined by edge $e={(j_e,k_e;D_e)}$, with $j_e=\min\{l:l\in (V_a \cup V_b) \setminus D_e\}$, $k_e=\max\{l:l\in ({\mathcal{V}_a} \cup {\mathcal{V}_b}) \setminus D_e\}$ and $D_e={\mathcal{V}_a} \cap {\mathcal{V}_b}$.

A vine copula is a graphical model describing the dependence of a $d$-variate random vector $\bm U = (U_1, \dots, U_d) \sim C$. The vine tree sequence is also called the vine \emph{structure} of the vine copula model. This model identifies each edge $e$ of the vine with a bivariate copula $c_{j_e, k_e ; D_e}$ (called \emph{pair-copula}). The joint density of the vine copula can then be written as the product of all pair-copula densities:
\begin{align}
c(\bm u) = \prod_{m=1}^{d-1} \prod_{e \in E_m} c_{j_e, k_e; D_e} \bigl(C_{j_e|D_e}(u_{j_e}|\bm u_{D_e}), C_{k_e|D_e}(u_{k_e}|\bm u_{D_e}) ; \, \bm u_{D_e} \bigr), \label{sec:vines:density_nonsimplified_eq}
\end{align}
where $\bm u_{D_e}:=(u_\ell)_{\ell \in D_e}$ is a subvector of $\bm u =(u_1, \dots, u_d) \in [0,1]^d$ and $C_{j_e|D_e}$ is the conditional distribution of $U_{j_e} | \bm U_{D_e} = \bm u_{D_e}$. The pair-copula density $c_{j_e, k_e; D_e}$ is the copula density corresponding to the two variables $U_{j_e}$ and $U_{k_e}$, conditional on $\bm U_{D_e} = \bm u_{D_e}$.

The density decomposition \eqref{sec:vines:density_nonsimplified_eq} holds for \emph{any} copula density $c$. In this general form, the pair-copulas $c_{j_e,k_e;D_e}$ depend on the value of the conditioning vector $\bm u_{D_e}$. To make the model more tractable, one usually makes the \emph{simplifying assumption} that the pair-copula densities do not change with $\bm u_{D_e}$. In this case, the model is called a simplified vine copula model and the corresponding density can be written as
\begin{align*}
c(\bm u) = \prod_{m=1}^{d-1} \prod_{e \in E_m} c_{j_e, k_e; D_e} \bigl(C_{j_e|D_e}(u_{j_e}|\bm u_{D_e}), C_{k_e|D_e}(u_{k_e}|\bm u_{D_e})\bigr).
%\label{sec:vines:density_simplified_eq}
\end{align*}

\begin{Example}
	The density of a simplified vine copula model corresponding to the tree sequence in \autoref{fig:trees} is
	\begin{align*}
	c(u_1, \dots, u_5) &= c_{1,2}(u_1, u_2) \times c_{1,3}(u_1,u_3) \times c_{3,4}(u_3,u_4) \times c_{3,5}(u_3,u_5)\\
	& \phantom{=} \times c_{2,3;1}(u_{2|1}, u_{3|1}) \times c_{1,4;3}(u_{1|3}, u_{4|3}) \times c_{1,5;3}(u_{1|3}, u_{5|3}) \\
	& \phantom{=} \times c_{2,4;1,3}(u_{2|1,3}, u_{4|1,3}) \times c_{4,5;1,3}(u_{4|1,3}, u_{5|1,3}) \\
	& \phantom{=} \times c_{2,5;1,3,4}(u_{2|1,3,4}, u_{5|1,3,4}),
	\end{align*}
	where we used the abbreviation $u_{j_e|D_e} := C_{j_e|D_e}(u_{j_e}|\bm u_{D_e})$.
\end{Example}

Vine copula densities involve conditional distributions $C_{j_e|D_e}$. We can express them in terms of conditional distributions corresponding to bivariate copulas in the previous tree as follows: Let $l_e \in D_e$ be another index such that $c_{j_e, l_e; D_e \setminus l_e}$ is a pair-copula in the previous tree, and define $D'_e= D _e \setminus l_e$. Then, we can express
\begin{align}
 C_{j_e|D_e}(u_{j_e}|\bm u_{D_e}) = h_{j_e|l_e;D'_e }\bigl(C_{j_e|D'_e}(u_{j_e}|\bm u_{D'_e})\, \big| \, C_{l_e|D'_e}(u_{l_e}|\bm u_{D'_e})\bigr), 
 \label{sec:vines:h_recursive_eq}
\end{align}
where the \emph{h-function} is defined as
\begin{align}
h_{j_e|l_e;D'_e} (u_{j_e} | u_{l_e}) := \int_{0}^{u_{j_e}} c_{j_e, l_e;D'_e}(v , u_{l_e}) dv = \frac{\partial C_{j_e, l_e; D'_e}(u_{j_e}, u_{l_e}|\bm u_{D'_e})}{\partial u_{l_e}}. \label{sec:vines:hfuncdef_eq}
\end{align}
The arguments $C_{j_e|D'_e}(u_{j_e}|\bm u_{D'_e})$ and $C_{l_e|D'_e}(u_{l_e}|\bm u_{D'_e})$ of the h-function in \eqref{sec:vines:h_recursive_eq} can be rewritten in the same manner. In each step of this recursion the conditioning set $D_e$ is reduced by one element. Eventually, this allows us to write any of the conditional distributions $C_{j_e|D_e}$ as a recursion over h-functions that are directly linked to the pair-copula densities in previous trees.
    \section{Nonparametric estimation of simplified vine copula models}
\label{sec:inference}

We now discuss how simplified vine copula models can be estimated nonparametrically.
First, we give an overview of nonparametric estimators of bivariate copula densities. Second, we outline a general step-wise estimation algorithm for the full vine copula density, which can be used with any bivariate copula density estimator. We also describe a data-driven structure selection algorithm that was initially proposed by \citet{Dissmann13}.

\subsection{Nonparametric estimation of bivariate copula densities}
\label{subsec:nonparametric}

The classical approach to density estimation is to assume a parametric model and estimate its parameters by maximum likelihood. There is a large variety of bivariate parametric copula models. Special classes are the elliptical copulas (including the Gaussian and Student t families), and the Archimedean class (including the Clayton, Frank and Gumbel families) \citep[for more, see,][]{Joe14}. However, parametric models notoriously lack flexibility and bear the risk of misspecification. Nonparametric density estimators  are designed to remedy these issues. In the context of copula densities, these estimators have to take the bounded support ($[0,1]^d$) into account. 

In the following we summarize the state-of-the-art of the major strands of nonparametric copula density estimation. For simplicity, we only consider the bivariate case. We assume throughout that we are given $n$ observations $(U_1^{(i)}, U_2^{(i)})$, $i = 1, \dots, n$, from a copula density $c$ that we want to estimate.

\subsubsection{Empirical Bernstein copula}\label{subsec:bern}
A classical tool in function approximation are Bernstein polynomials (see \citet{Lorentz:53}). The normalized Bernstein polynomial of degree $K$ is defined as 
\begin{align*}
B_{Kk}(u)=(K+1) \binom{K}{k} (u)^k (1-u)^{m-k}, \quad \mbox{for }  k=0,\dots,K.
\end{align*}
The collection of all Bernstein polynomials form a basis of the space of all square-integrable functions on $[0,1]$. A natural idea is to approximate an arbitrary function by a linear combination of a finite number of basis functions. Based on this idea, \citet{Sancetta04} defined the Bernstein copula density. It is an approximation of the true copula density, and can be expressed as
\begin{equation*}
\widetilde c(u_1,u_2)= \sum_{k_1=0}^{K}\sum_{k_2=0}^{K}B_{Kk_1}(u_1) B_{Kk_2}(u_2)  v_{k_1, k_2} ,
\end{equation*}
where
\begin{align*}
v_{k_1,k_2} = \int_{k_1/\bar K}^{(k_1 + 1)/\bar K} \int_{k_2/\bar K}^{(k_2 + 1)/\bar K} c(u_1, u_2) du_1 du_2.
\end{align*}
and $\bar K = (K +1)$. Note that the coefficient $v_{k_1,k_2}$ describes the probability that $(U_1^{(i)}, U_2^{(i)})$ is contained in the cell $[k_1/\bar K , (k_1 + 1)/\bar K ] \times [k_2/\bar K , (k_2 + 1)/\bar K ]$. The empirical copula density estimator is defined by $\widetilde c(u_1,u_2)$, but replacing $v_{k_1,k_2}$ by the empirical frequencies obtained from a contingency table:
\begin{equation*}
\wh c(u_1,u_2)=  \sum_{k_1=0}^{K}\sum_{k_2=0}^{K}B_{Kk_1}(u_1) B_{Kk_2}(u_2) \wh  v_{k_1, k_2},
\end{equation*}
where 
\begin{align*}
\wh v_{k_1,k_2} = \frac{1}{n} \times \#\bigl\{(U_1^{(i)}, U_2^{(i)}) \in [k_1/\bar K , (k_1 + 1)/\bar K ] \times [k_2/\bar K , (k_2 + 1)/\bar K ]\bigr\},
\end{align*}
which is the maximum-likelihood estimator for $v_{k_1,k_2}$.

The Bernstein copula density estimator was used in the context of vine copulas by \citet{SchefferWeiß16}. As the marginal distributions of the Bernstein copula density do not need to be uniform, the authors calculate an approximation to the contingency table by solving a quadratic program, imposing constraints for uniform marginal distributions. The smoothing parameter for the Bernstein copula density estimator is $K$, the number of knots. \citet{Rose2015} proposed selection rules for $K$ that adapt to the sample size and strength of dependence. Our implementation is available in the \texttt{kdecopula} R package \citep{kdecopula}, and uses the rule
\begin{align*}
K^{opt} = \lfloor n^{1/3}  \exp(\vert\wh \rho\vert^{1/n})  (\vert \wh \rho\vert + 0.1)\rfloor,
\end{align*}
where $\wh \rho$ is the empirical Spearman's $\rho$.

\subsubsection{Penalized Bernstein polynomials and B-splines}
\label{subsec:splines}

For fixed $K$, the Bernstein copula density estimator is a parametric model with $(K+1)^2$ parameters. As any parametric model with many parameters, it is prone to overfitting. \citet{ScheKau:12b} formulated a penalized likelihood approach to gain control of the smoothness of the fit. 

Viewing the Bernstein copula density as a parametric model with parameter vector $\bv=(v_{00},\dots,v_{0K},\dots,v_{KK})$, i.e.,
\begin{eqnarray}
\widetilde{c}(u_1,u_2;\bv)=\sum_{k_1=0}^K \sum_{k_2=0}^K B_{K{k_1}}(u_1) B_{K{k_2}}(u_2) v_{k_1,k_2},\label{eq:splinerep}
\end{eqnarray}
we can estimate the parameters by maximizing the log-likelihood,
\begin{align} \label{eq:loglik}
\ell(\bv) = \log \sum_{i=1}^n \widetilde{c}\bigl(U_1^{(i)}, U_2^{(i)};\bv\bigr).
\end{align}
As each of the normalized Bernstein polynomials is a density, the weighted sum of normalized Bernstein polynomials is a density, if we ensure that
\begin{equation}\label{eq:constraint0}
\sum_{k_1,k_2}v_{k_1,k_2}=1, \qquad v_{k_1,k_2}\geq 0.
\end{equation}
We will need additional constraints to enforce uniform marginal distributions: for Bernstein polynomials $\int \tilde{c}(u_1,u_2) \text{ d}u_{1} \equiv 1$ holds if the marginal coefficients fulfill
\begin{equation}\label{eq:constraint3}
 v_{k_{1} .} = \sum_{k_2} v_{k_1,k_2}=1/(K+1), \quad \mbox{for all } k_1=0,\dots,K.
\end{equation}
The same constraints follow for $\int \tilde{c}(u_1,u_2) \text{ d}u_{2} \equiv 1$. These constraints can be formulated in matrix notation yielding
\begin{equation}\label{eq:constraint1}
 A^T_{K}{\boldsymbol v}={\boldsymbol 1/(K+1)}
\end{equation}
where $A_{K}$ sums up the elements of $v_{k_1,k_2}$ column-wise (i.e. over $k_2$) and row-wise (i.e. over $k_1$), i.e. $A_K^T=((I_K \otimes {\boldsymbol 1}_K^{T})),({\boldsymbol 1}_K^T \otimes I_K))$, where ${\boldsymbol 1}_K$ is the column vector of dimension $K$ with elements 1 and $I_K$ is the $K$ dimensional identity matrix.\par

The log-likelihood \eqref{eq:loglik} can be maximized under these constraints using quadratic programming \citep[e.g., with the {\ttfamily quadprog} {\ttfamily R} package][]{quadprog}. But since this is a parametric model with many parameters, the fitted copula density may be wiggly \citep[see e.g.,][]{wahba:90}. This issue can be resolved by imposing an appropriate penalty on the basis coefficients. \citet{ScheKau:12b} postulate that the integrated squared second order derivatives are small \citep[see also,][]{Wood:06} and formulate the penalty as
\begin{equation*}%\label{eq:penalty}
 \int \left(\frac{\partial^2 \widetilde{c}(u_1,u_2;\bv)}{(\partial u_{1})^2}\right)^2+\left(\frac{\partial^2\widetilde{c}(u_1,u_2;\bv)}{(\partial u_2)^2} \right)^2 \text{ d}u_1 \text{ d}u_2.
\end{equation*}
This can be written as a quadratic form of a penalty matrix $\mathbf{P}$ \citep[see,][Appendix]{ScheKau:12b} and the corresponding penalized log-likelihood is defined as
\begin{eqnarray}\label{eq:penloglik}
 \ell^p(\bv,\lshort)= \ell(\bv)-\frac12\lshort{\bv}^T \mathbf{P} \bv.
\end{eqnarray}
The penalty parameter $\lshort$ needs to be selected adequately, that is data driven. In Section 2.5 of \citet{ScheKau:12b}, the authors propose a method that formulates the penalized likelihood approach as linear mixed model and comprehend the penalty as normal prior imposed on the coefficient vector. We apply this methodology, too.\par
Additionally, \citet{ScheKau:12b} propose to use B-spline basis functions (see \citet{Boor:78}) instead of Bernstein polynomials in \eqref{eq:splinerep}. They replace each $B_{Kk}$ in \eqref{eq:splinerep}  with a B-spline, located at equidistant knots $\kappa_k=k/K$ with $k=0,\dots,K-1+q$, normalized so that it satisfies $\int_0^1 B_{Kk}(u) \text{ d}u=1$ for $k=0,\dots,K$. \citet{ScheKau:12b} only used normalized linear $(q=1)$ B-splines. To allow for more flexibility, we will also use normalized quadratic $(q=2)$ B-splines in our study.\par
In order to guarantee that $\widetilde c(u_1,u_2; \bv)$ is a bivariate copula density, we impose similar constraints as the ones for the Bernstein polynomials. The linear constraints \eqref{eq:constraint0} will be the same for B-splines, but the uniform margins condition \eqref{eq:constraint3} has to be adapted. The condition takes the form $A_{K}\bv={\boldsymbol 1}$ with $A_{K}=\boldsymbol{B}_K(\kappa)$, choosing
\begin{align*}
 \kappa= \begin{cases}
\kappa_0,\dots,\kappa_{K} , &\mbox{for linear B-splines},\\
0,\frac{\kappa_1-\kappa_0}{2}+\kappa_0,\frac{\kappa_2-\kappa_1}{2}+\kappa_1,\dots,\frac{\kappa_{K+1}-\kappa_{K}}{2}+\kappa_{K},1, &\mbox{for quadratic B-splines.}
\end{cases} 
\end{align*}

For the penalization, we work with a penalty on the $m$-th order differences of the spline coefficients $\bv$, as suggested for B-spline smoothing in \citet{EilMar:96}, defining a penalty matrix $\mathbf{P^m}$, where we choose $m=q+1$. Further details of this smoothing concept can be found in \citet{Ruppert-etal:03}. In the following, we define the difference based penalty matrix $\mathbf{P^m}$ for the $m$-order differences through
\begin{equation}\label{eq:penalty2}
 \mathbf{P^m} :=({\mathbf 1}_{K+1}\otimes L_m)^T (L_m \otimes {\mathbf 1}_{K+1})
\end{equation}
with e.g.
\begin{equation*}L_1=
\begin{pmatrix}
1 & -1 & 0 & \cdots & 0\\
0 &  1 &-1 & \ddots & \vdots \\
\vdots & \ddots &\ddots & \ddots & 0 \\
0 & \cdots &0 &1&-1
\end{pmatrix}.
\end{equation*}
Then for B-splines, the penalized log-likelihood becomes
\begin{eqnarray}\label{eq:penloglik2}
 l^p(\bv,\lshort)= l(\bv)-\frac12\lshort{\bv}^T \mathbf{P^m} \bv.
\end{eqnarray}
Note, that we achieve an independence copula, if we set the penalty parameter $\lambda$ to infinity in \eqref{eq:penloglik} or \eqref{eq:penloglik2}. The penalized Bernstein and B-splines estimators are implemented in the R package \texttt{penRvine} \citep{penRvine}. The package will be published soon and is available upon request.
%---------------------------------------------------------------------------------------%

\subsubsection{Kernel weighted local likelihood}
\label{subsec:kernel}

Kernel estimators are well-established tools for nonparametric density estimation. Several kernel methods have been tailored to the problem of copula density estimation. Their main challenge is to avoid bias and consistency issues at the boundaries of the support. The earliest contribution is the mirror-reflection method \citep{Gijbels90}. Later, \citet{Charpentier06} extended the beta kernel density estimator of \citet{Chen93} to the bivariate case. 

The more recent contributions all focus on a transformation trick. Assume we want to estimate a copula density $c$ given a random sample $\bigl(U_1^{(i)}, U_2^{(i)}\bigr), i = 1, \dots, n$. Let $\Phi$ be the standard normal distribution function and $\phi$ its density. Then the random vectors $(Z_1^{(i)}, Z_2^{(i)}) = \bigl(\Phi^{-1}(U_1^{(i)}), \Phi^{-1}(U_2^{(i)})\bigr)$ have normally distributed margins and are supported on the full $\R^2$. In this domain, kernel density estimators work very well and do not suffer from any boundary problems. By Sklar's Theorem for densities, eq.\ \eqref{sec:vines:sklar_density_eq}, the density $f$ of $(Z_1^{(i)}, Z_2^{(i)})$ decomposes to
\begin{align}
f(z_1, z_2) = c\bigl(\Phi(z_1), \Phi(z_2)\bigr) \phi(z_1) \phi(z_2), \qquad \mbox{for all } (z_1, z_2) \in \R.
\label{subsec:nonparametric:trafo_eq}
\end{align}
By isolating $c$ in \eqref{subsec:nonparametric:trafo_eq} and the change of variables $u_j = \Phi(z_j), j = 1, 2$, we get
\begin{align}
	c(u_1, u_2) = \frac{f\bigl(\Phi^{-1}(u_1), \Phi^{-1}(u_2)\bigr)}{\phi\bigl(\Phi^{-1}(u_1)\bigr)\phi\bigl(\Phi^{-1}(u_2)\bigr)}.
    \label{subsec:nonparametric:trafo2_eq}
\end{align}
We can use any kernel estimator $\wh f$ of $f$ to define an kernel estimator of the copula density $c$ via \eqref{subsec:nonparametric:trafo2_eq}: 
\begin{align}
	\wh c(u_1, u_2) = \frac{\wh f\bigl(\Phi^{-1}(u_1), \Phi^{-1}(u_2)\bigr)}{\phi\bigl(\Phi^{-1}(u_1)\bigr)\phi\bigl(\Phi^{-1}(u_2)\bigr)}.
    \label{subsec:nonparametric:trafo3_eq}
\end{align}
Estimators of this kind have an interesting feature. The denominator of \eqref{subsec:nonparametric:trafo3_eq} vanishes when $u_1$ or $u_2$ tend to zero or one. If the numerator vanishes at a slower rate, the estimated copula density explodes towards the corners of the unit square. This behavior is common for many popular parametric families, including the Gauss, Student, Gumbel, and Clayton families. The transformation estimator \eqref{subsec:nonparametric:trafo3_eq} is well suited to resemble such shapes. However, its variance will also explode towards the corners and the estimator will be numerically unstable. To accommodate for this, we restrict the estimator to $[0.001, 0.999]^2$ and set estimates outside of this region to the closest properly defined estimate.

To estimate the density $f$, \citet{Nagler15} proposed to use the classical bivariate kernel density estimator \citep[see, e.g.,][]{Silverman86}. We will extend this approach by resorting to the more general class of local polynomial likelihood estimators; see \citet{Loader99} for a general account and \citet{Geenens14a} in the context of bivariate copula estimation.

Assume that the log-density $\log f(z_1,z_2)$ of the random vector $\bm Z^{(i)} = (Z_1^{(i)}, Z_2^{(i)})$ can be approximated locally by a polynomial of order $q$. For example, using a log-quadratic expansion, we get 
\begin{align*}
\log f(z_1',z_2') &\approx P_{\bm a}(\bm z) \\
&=a_1+ a_2 (z_1-z_1') + a_3(z_2-z_2')  \\
&\phantom{=} + a_4(z_1-z_1')^2 +  a_5(z_1-z_1')(z_2-z_2') + a_6(z_2-z_2')^2
\end{align*}
for $(z_1',z_2')$ in the neighborhood of $\bm z = (z_1,z_2)$. The polynomial coefficients $\bm a$ can be found by solving the weighted maximum likelihood problem
\begin{align*}
\widehat{\bm a} &= \arg\max_{\bm a \in \R^6} \biggl\{\sum_{i=1}^{n} \bm K\bigl( B^{-1} (\bm z - \bm Z^{(i)}) \bigr) P_{\bm a}(\bm z - \bm Z^{(i)}) \\
&\phantom{= \arg\max_{\bm a \in \R^6} \biggl\{} - n \int_{\R^2}\bm K \bigl( B^{-1} (\bm z - \bm s) \bigr)\exp\bigl(P_{\bm a}(\bm z - \bm s)  d\bm s \biggr\},
\end{align*}
where the kernel $K$ is a symmetric probability density function,\linebreak $\bm K(\bm z) = K(z_1) K(z_2)$ is the product kernel, and $B \in \R^{2\times 2}$ is a matrix with $\det(B) > 0$. $B$ is called the bandwidth matrix and controls the degree of smoothing. The kernel $K$ serves as a weight function that localizes the above optimization problem around $\bm z$. 

We obtain $\wh a_1$ as an estimate for $\log f(z_1,z_2)$ and, consequently, $\exp(\wh a_1)$ as an estimate for $f(z_1,z_2)$. An estimate of the copula density can be obtained by plugging this estimate in \eqref{subsec:nonparametric:trafo2_eq}. For a detailed treatment of this estimator's asymptotic behavior we refer to \citet{Geenens14a}. In general, the estimator does not yield a \emph{bona fide} copula density because the margins may not be uniform. This issue can be resolved by normalizing the density estimate \citep[see,][for details]{Nagler16}.

For applications of the estimator, an appropriate choice of the bandwidth matrix is crucial. For the local constant approximation, a simple rule of thumb was shown to perform well in \citet{Nagler16}. We use an improved version of this rule that also adjusts to the degree of the polynomial $q$:
\begin{align*}
B_{\mathrm{rot}} = \nu_q n^{-1/(4q^* + 2)} \widehat{\Sigma}_{\bm Z}^{1/2}, \quad q^* = 1 + \lfloor q/2 \rfloor,
\end{align*}
where $\widehat{\Sigma}_{\bm Z}$ is the empirical covariance matrix of $\bm Z^{(i)}$, $i = 1, \dots, n$, and $\nu_0 = 1.25$, $\nu_1 = \nu_2 = 5$. An implementation of the estimator is available in the R package \texttt{kdecopula} \citep{kdecopula}.

%---------------------------------------------------------------------------------------%

\subsection{Step-wise estimation of vine copula densities}
\label{subsec:stepwise}

\begin{algorithm}[t]
	\caption{Sequential estimation of simplified vine copula densities}
	\label{subsec:stepwise:seqest_alg}
	{\bfseries Input:} Observations $(U_1^{(i)}, \dots, U_d^{(i)})$, $i=1, \dots, n$, vine structure $(E_1, \dots, E_{d-1})$. \\
	{\bfseries Output:} Estimates of pair-copula densities and h-functions required to evaluate the vine copula density \eqref{subsec:stepwise:density_simplified_eq}.\\	---------------------------------------------------------------------------------------------------------\\
	{\bfseries for} $m=1, \dots, d-1$:\\
	\hspace*{2em} {\bfseries for all} $e \in E_m$:\\[-12pt]
	\hspace*{1em}
	\begin{minipage}[t]{0.9\textwidth}
		\begin{enumerate}
			\item Based on $\bigl(\wh U_{j_e|D_e}^{(i)}, \wh U_{k_e|D_e}^{(i)}\bigr)_{i=1,\dots,n}$, obtain an estimate of the copula density $c_{j_e, k_e;D_e}$  which we denote as $\wh c_{j_e, k_e;D_e}$. 
            \item Derive corresponding estimates of the h-functions  $\wh h_{j_e| k_e;D_e}$, $\wh h_{k_e| j_e;D_e}$ by integration (eq.\ \eqref{sec:vines:hfuncdef_eq}).
			\item Set
			\begin{align*}
			\wh U_{j_e|D_e \cup k_e}^{(i)} &:= \wh h_{j_e|k_e ; D_e}\bigl(\wh U_{j_e|D_e}^{(i)}\big| \wh{ U}_{k_e | D_e}^{(i)}\bigr), \\
			\wh U_{k_e|D_e \cup j_e}^{(i)} &:= \wh h_{k_e|j_e ; D_e}\bigl(\wh U_{k_e|D_e}^{(i)}\big| \wh{U}_{ j_e|D_e}^{(i)}\bigr), \quad i = 1, \dots, n.
			\end{align*}
		\end{enumerate}
	\end{minipage}
	\hspace*{2em} {\bfseries end for}\\
	{\bfseries end for}
\end{algorithm}

We now turn to the question how a simplified vine copula density can be estimated. Most commonly, this is done in a sequential procedure introduced by \citet{Aas09}. The procedure is generic in the sense that it can be used with any consistent estimator for a bivariate copula. It is summarized in \autoref{subsec:stepwise:seqest_alg}.

From now on we use $c$ to denote a $d$-dimensional vine copula density. Assume we have a random sample $\bm U^{(i)}=\bigl(U_1^{(i)}, \dots, U_d^{(i)}\bigr), i = 1, \dots, n,$ from $c$. Recall that this density can be written as
\begin{align}
c(\bm u) = \prod_{m=1}^{d-1} \prod_{e \in E_m} c_{j_e, k_e; D_e} (C_{j_e|D_e}(u_{j_e}|\bm u_{D_e}), C_{k_e|D_e}(u_{k_e}|\bm u_{D_e})). \label{subsec:stepwise:density_simplified_eq}
\end{align}
In the first tree, the conditioning set $D_e$ is empty. So for $e \in E_1$, estimation of the pair-copula densities $c_{j_e,k_e;D_e}$ is straightforward, since no conditioning is involved. We simply apply one of the estimators from \autoref{subsec:nonparametric} to the bivariate random vectors $(U_{j_e}^{(i)}, U_{k_e}^{(i)})$. This gives us estimates $\wh c_{j_e, k_e;D_e}, e \in E_1$. By one-dimensional integration (eq.\ \eqref{sec:vines:hfuncdef_eq}) we can derive estimates of the corresponding h-functions. They can be derived in closed form  for Bernstein and B-spline estimators \citep[see,][]{ScheKau:12b}. For kernel estimators, the h-functions have to be computed numerically.\par
In a next step, we transform the initial copula data by applying the estimated h-functions to obtain pseudo-observations from the pair-copulas in the second tree. Using these, we can estimate the pair-copula densities  $c_{j_e,k_e;D_e}$, $e \in E_2$. We iterate through the trees in this manner until all pair-copula densities and h-functions have been estimated.\par
\citet[Theorem 1]{Nagler15} show that simplified vine copula density estimators defined by \autoref{subsec:stepwise:seqest_alg} are consistent under rather mild conditions. An appealing property of these estimators is the absence of curse of dimensionality: the convergence rate does not depend on the dimension. In fact, the convergence rate achieves the optimal rate for a two-dimensional nonparametric density estimator.

%---------------------------------------------------------------------------------------%

\subsection{Selection strategies for the vine structure}
\label{subsec:structure}

\begin{algorithm}[t]
	\caption{Sequential estimation and structure selection for simplified vine copula models}
	\label{subsec:stepwise:structure_alg}
	{\bfseries Input:} Observations $(U_1^{(i)}, \dots, U_d^{(i)})$, $i=1, \dots, n$. \\
	{\bfseries Output:} Vine structure $(E_1, \dots, E_{d-1})$ and estimates of pair-copula densities and h-functions required to evaluate the vine copula density \eqref{subsec:stepwise:density_simplified_eq}.\\	
    ---------------------------------------------------------------------------------------------------------\\
	{\bfseries for} $m=1, \dots, d-1$:\\
   \hspace*{2em} 
   \begin{minipage}[t]{0.9\textwidth}
  Calculate weights $w_e$ for all possible edges $e = \{j_e,k_e;D_e\}$ that satisfy the proximity condition (see \autoref{sec:vines}) and select the edge set $E_m$ as
    \begin{align*}
	E_m = \underset{E^*_m}{\arg\max} \sum_{e \in E^*_m} w_{e},
	\end{align*}
    under the constraint that ${E^*_m}$ corresponds to a spanning tree.
\end{minipage}
  
	\hspace*{2em} {\bfseries for all} $e \in E_m$:\\[-12pt]
	\hspace*{1em}
	\begin{minipage}[t]{0.9\textwidth}
		\begin{enumerate}
			\item Based on $\bigl(\wh U_{j_e|D_e}^{(i)}, \wh U_{k_e|D_e}^{(i)}\bigr)_{i=1,\dots,n}$, obtain an estimate of the copula density $c_{j_e, k_e;D_e}$  which we denote as $\wh c_{j_e, k_e;D_e}$. 
            \item Derive corresponding estimates of the h-functions  $\wh h_{j_e| k_e;D_e}$, $\wh h_{k_e| j_e;D_e}$ by integration (eq.\ \eqref{sec:vines:hfuncdef_eq}).
			\item Set
			\begin{align*}
			\wh U_{j_e|D_e \cup k_e}^{(i)} &:= \wh h_{j_e|k_e ; D_e}\bigl(\wh U_{j_e|D_e}^{(i)}\big| \wh{ U}_{k_e | D_e}^{(i)}\bigr), \\
			\wh U_{k_e|D_e \cup j_e}^{(i)} &:= \wh h_{k_e|j_e ; D_e}\bigl(\wh U_{k_e|D_e}^{(i)}\big| \wh{U}_{ j_e|D_e}^{(i)}\bigr), \quad i = 1, \dots, n.
			\end{align*}
		\end{enumerate}
	\end{minipage}
	\hspace*{2em} {\bfseries end for}\\
	{\bfseries end for}
\end{algorithm}
So far we assumed that the structure of the vine (i.e., the edge sets $E_1, \dots, E_{d-1}$) is known. In practice, however, the structure has to be chosen by the statistician. This choice is very difficult, since there are $d!/2 \times d^{(d - 2)(d-3)/2}$ possible vine structures \citep{Morales10}, which grows excessively with $d$. When $d$ is very small, it may still be practicable to estimate vine copula models for all possible structures and compare them by a suitable criterion (such as AIC). But already for a moderate number of dimensions one has to rely on heuristics.\par
\citet{Dissmann13} proposed a selection algorithm that seeks to capture most of the dependence in the first couple of trees. This is achieved by finding the maximum spanning tree using a dependence measure as edge weights, e.g., the absolute value of the empirical Kendall's $\tau$. The resulting estimation and structure selection procedure is summarized in a general form in \autoref{subsec:stepwise:structure_alg}.\par
Several specifications of the edge weight were investigated in a fully parametric context by \citet{Czado13}. The most common edge weight $w_e$ is the absolute value of the empirical Kendall's $\tau$ as proposed by \citet{Dissmann13} and used in a non-parametric context by \citet{Nagler15}. \citet{ScheKau:12b}, on the other hand, used the corrected Akaike information criterion (cAIC) \citep{Hurvich98} as edge weight. Using the notation of  Algorithm 2, the cAIC for edge $e$ is defined as 
\begin{equation} \label{eq:AIC}
 \text{cAIC}_e =-2 \ell_e  + 2 \text{df}_e+ \frac{2 \text{df}_e (\text{df}_e +1)}{n- \text{df}_e-1},
\end{equation}
where 
\begin{align*}
\ell_e = \sum_{i = 1}^n \ln \wh c_{j_e, k_e;D_e}\bigl(\wh U_{j_e|D_e}^{(i)}, \wh U_{k_e|D_e}^{(i)}\bigr),
\end{align*}
is the log-likelihood and $\mathrm{df}_e$ is the \emph{effective degrees of freedom (EDF)} of the estimator $\wh c_e$.
For explicit formulas for the EDF we refer to \citet{ScheKau:12b}  for the spline approach and to \citet[Section 5.3.2]{Loader99} for the kernel estimators. For parametric copula estimation, the EDF equals the number of estimated parameters for the chosen copula family.\par
From a computational point of view, the cAIC has a big disadvantage: before a tree can be selected, the pair-copulas of all possible edges in this tree have to be estimated. Just for the first tree, this amounts to estimating $\genfrac{(}{)}{0pt}{}{d}{2}$ bivariate copula densities. The empirical Kendall's $\tau$ on the other hand can be computed rapidly for all pairs. It allows to select the tree structure before any pair-copula has been estimated. Then, only $d-1$ pair-copulas have to be estimated in the first tree. The situation is similar for subsequent trees. Both approaches will be compared in our simulation study with regard to estimation accuracy and speed.
    \section{Description of the simulation study design} 
\label{sec:simulations}
We compare the performance of the vine copula density estimators discussed in \autoref{sec:inference} over a wide range of  scenarios. We consider several specifications of sample size, dimension, strength of dependence, and tail dependence. We randomize the simulation models and characterize the scenarios by probability distributions for the pair-copula families and dependence parameters. A detailed description of the study design procedure will be given in the following sections. 
\subsection{Simulation scenarios based on model randomization}
To investigate how various factors influence the estimators' performance, we create a number of scenarios. Each of these scenarios is characterized by a combination of the  factors shown in \autoref{sec:simulations:factors_tab}.

\begin{table}[t]
	\begin{tabular}{c|c|c|c}
		Dimension $d$ & Sample Size $n$ & Type of dependence & Strength of dependence \\ \hline
        5 & 400 & only tail dependence & weak \\ 
        10 & 2\,000& no tail dependence & strong \\
         & & both types & \\
	\end{tabular}
    \caption{List of factors that determine the set of simulation scenarios.}
    \label{sec:simulations:factors_tab}
\end{table}

\begin{figure}
\centering
\includegraphics[width = 0.8\textwidth]{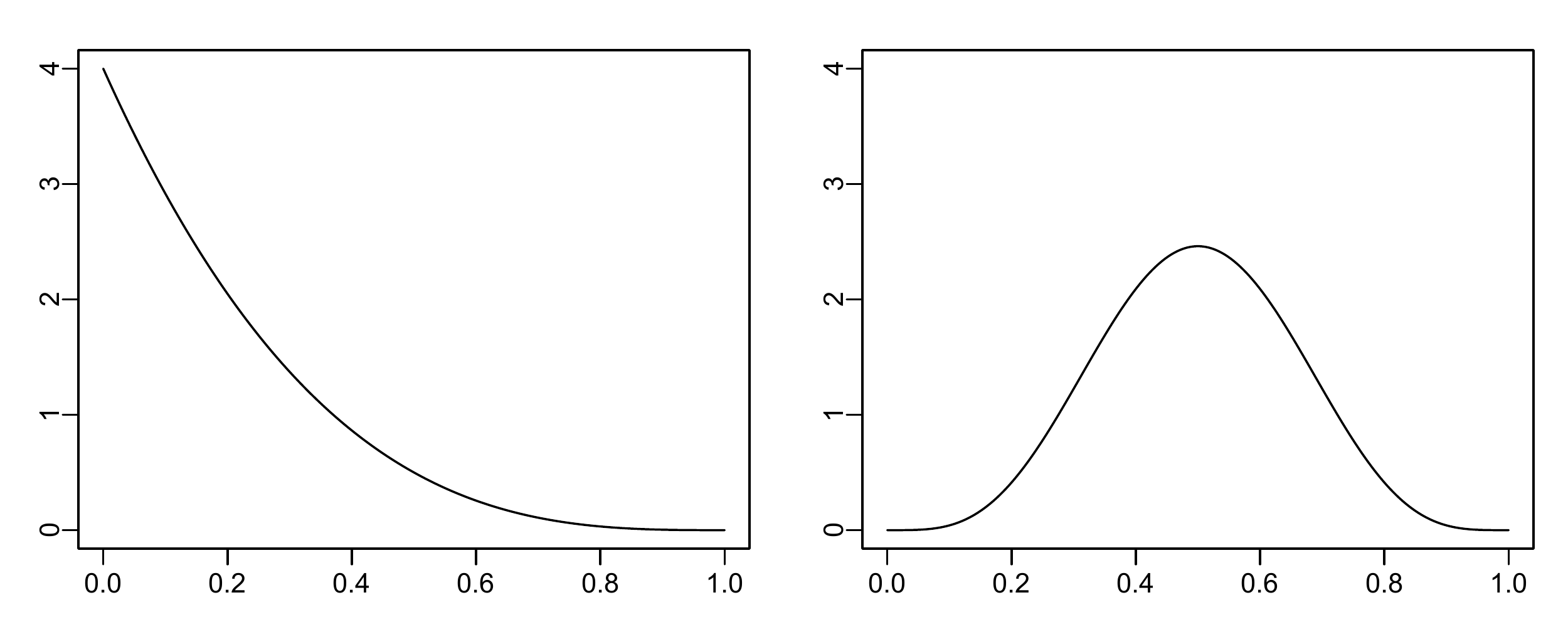}\caption{Densities for the simulation of absolute Kendall's $\tau$ in the scenarios with weak (left) and strong (right) dependence.}
\label{beta}
\end{figure}

To make the results for a particular dependence scenario as general as possible, we randomly generate a model in the following steps:
	\begin{enumerate}[Step 1.]
		\item \textbf{Draw R-vine structure:}\\
		We do this in the following steps:
        \begin{enumerate}[(i)]
        \item Draw $n$ samples for $d$ independent uniform random variables, $\tilde U_{i, j}$, $i = 1, \dots, n$, $j = 1, \dots, d$.
        \item On these samples, run the structure selection algorithm of \citet{Dissmann13} (only allowing for the independence family).
        \item Set the model structure to the one selected by the algorithm.
        \end{enumerate}  
        
		\item \textbf{Draw pair-copula families:}	
		\begin{itemize}
		\item \emph{only tail dependent copulas}: draw each of the $d(d-1)/2$ pair-copula families with equal probabilities from the Student t- ($df=4$), Gumbel (with rotations) and Clayton (with rotations) copulas.
        
		\item \emph{no tail dependence}: draw each of the $d(d-1)/2$ pair-copula families with equal probabilities from  the Gaussian and Frank copulas.
        
		\item \emph{both}: for each of $d(d-1)/2$ pair-copulas:
		\begin{enumerate}[(i)]
		\item choose with equal probabilities whether the copula has tail dependence or not,
		\item proceed as above.
\end{enumerate}		 
\end{itemize} 

	\item \textbf{Draw pair-copula parameters:}\\
	For each pair-copula:
	\begin{enumerate}[(i)]
	\item Randomly generate the absolute value of Kendall's $\tau$ from the following distributions:
	\begin{itemize}
	\item \emph{weak dependence:} $Beta(1,4)$-distribution ($E[\vert \tau \vert] = 0.2$),
	\item \emph{strong dependence:} $Beta(5,5)$-distribution ($E[\vert \tau \vert] = 0.5$).
	\end{itemize}
	The densities are shown in \autoref{beta}.
	\item Randomly choose the sign of Kendall's $\tau$ as $Bernoulli(0.5)$ variable.
    \item Usually, partial dependence is weaker than direct pair-wise dependence. To mimic this behavior we decrease the simulated absolute Kendall's $\tau$ by a factor of $0.8^m$, where $m$ is the tree level of the pair-copula. 
	\item For all families under consideration there is a one-to-one relationship between the copula parameter and Kendall's $\tau$ \citep[see, e.g.,][Table 2]{Brechmann13}. Hence, we set the copula parameter by inversion of the reduced value of Kendall's $\tau$. 
\end{enumerate}		

	\item \textbf{Draw observations from the final model:} \\
    With the selected structure, copula families and their parameters, the vine copula model is fully specified. We can draw random samples from this vine copula model using the algorithm of \citet{Stoeber12} \citep[as implemented in the \texttt{VineCopula} R library][]{VineCopula}.
\end{enumerate}

The stochastic model characterized by steps 1--4 can be interpreted as a whole. It is a mixture of vine copula models, mixed over its structure, families, and parameter. The mixing distribution for the families is uniform over sets determined by the `type of dependence' hyper-parameter. The mixing distribution for the absolute Kendall's $\tau$ follows a $\mathrm{Beta}$ distribution with parameters characterized by the `strength of dependence` hyper-parameter. Each scenario corresponds to a particular specification of the mixture's hyperparameters. The benefit of this construction is that it yields models that are representative for a wide range of scenarios encountered in practice. It also limits the degrees of freedom we would have when specifying all pair-copula families and parameters manually.

\subsection{Estimation methods}

We compare the following pair-copula estimators:
\begin{itemize}
	\item \texttt{par}: parametric estimator as implemented in the function \texttt{BiCopSelect} of the R package \texttt{VineCopula} \citep{VineCopula}. It estimates the parameters for several parametric families and selects the best model based on AIC. The implemented families are: Independence, Gaussian, Student t, Clayton, Frank, Gumbel, Joe, BB1, BB6, BB7, BB8, Tawn types I and II,
    \item \texttt{bern}: non-penalized Bernstein estimator (see \autoref{subsec:bern}),
    \item \texttt{pbern}: penalized Bernstein estimator (see \autoref{subsec:bern}) with $K=14$ knots,
    \item \texttt{pspl1}: penalized linear B-spline estimator (see \autoref{subsec:splines}) with $K=14$,
    \item \texttt{pspl2}: penalized quadratic B-spline estimator (see \autoref{subsec:splines}) with $K=10$,
    \item \texttt{tll0}: transformation local likelihood kernel estimator of degree $q=0$ (see \autoref{subsec:kernel}),
    \item \texttt{tll1}: transformation local likelihood kernel estimator of degree $q=1$ (see \autoref{subsec:kernel}),
    \item \texttt{tll2}: transformation local likelihood kernel estimator of degree $q=2$ (see \autoref{subsec:kernel}).
\end{itemize}
We further implemented two structure selection methods for each estimation pair-copula estimator (based on Kendall's $\tau$ and cAIC, see \autoref{subsec:structure}); additionally we computed each estimator under the true structure. 

\subsection{Performance measurement}

As a performance measure, we choose the \emph{integrated absolute error (IAE)}
\begin{align*}
\mathrm{IAE} = \int_{[0,1]^d} \vert \wh c(\bm u) - c(\bm u) \vert d\bm u,
\end{align*}
where $\wh c$ is the estimated and $c$ is the true copula density. The above expression requires us to calculate a $d$-dimensional integral, which can be difficult when $d$ becomes large. To overcome this, we estimate this integral via importance sampling Monte Carlo integration \citep[e.g.,][Section 5.2]{Ripley87}. That is,
\begin{align*}
\wh{ \mathrm{IAE}} = \bm{\frac{1}{N}} \sum_{i=1, \dots, N} \frac{\vert \wh c(\bm U_i) - c(\bm U_i)\vert}{c(\bm U_i) },
\end{align*}
where $\bm U_i \stackrel{iid}{\sim} c$ is a random vector drawn from the true copula density $c$. This results in an unbiased estimator of the IAE with relatively small variance: usually the numerator is large/small when the denominator is large/small. Hence, the variance of the terms of the sum is small and, thereby, the variance of the sum is small. All results will be based on an importance sample of size $N = 1\,000$.\par
For each estimator and  each possible simulation scenario emerging from \autoref{sec:simulations:factors_tab}, we record the $\widehat{\mathrm{IAE}}$ on $R= 100$ simulated data sets.
    \section{Results}
\label{sec:results}
\begin{figure}
\centering
\includegraphics[width = \textwidth]{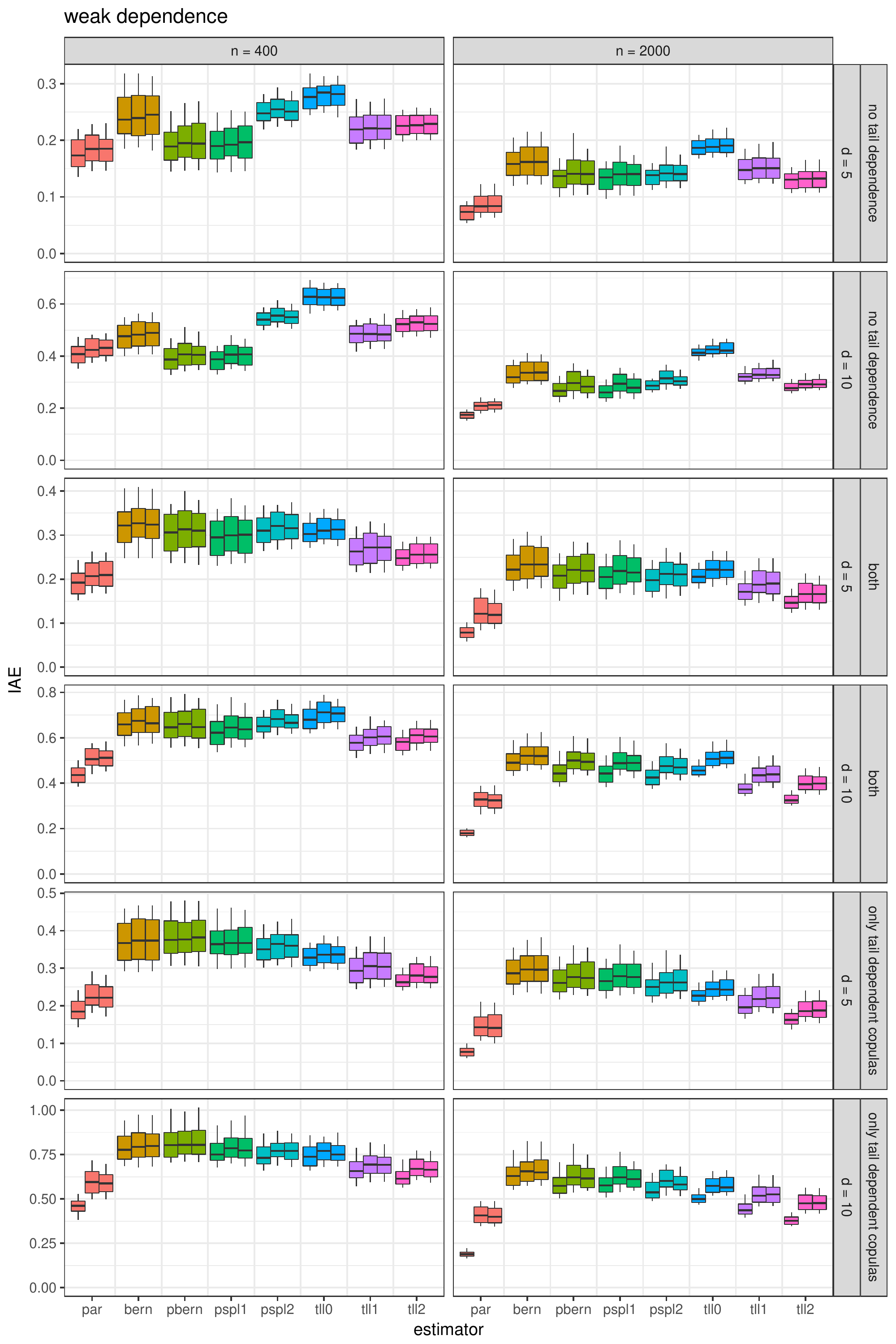}
\caption{\emph{weak dependence:} the box plots show the IAE achieved by each estimation method. Results are split by sample size, dimension, and type of dependence. Per estimator there are three boxes, corresponding to estimation under known structure, selection by Kendall's $\tau$, and selection by cAIC (from left to right).}
\label{fig:weak}
\end{figure}

\begin{figure}
\centering
\includegraphics[width = \textwidth]{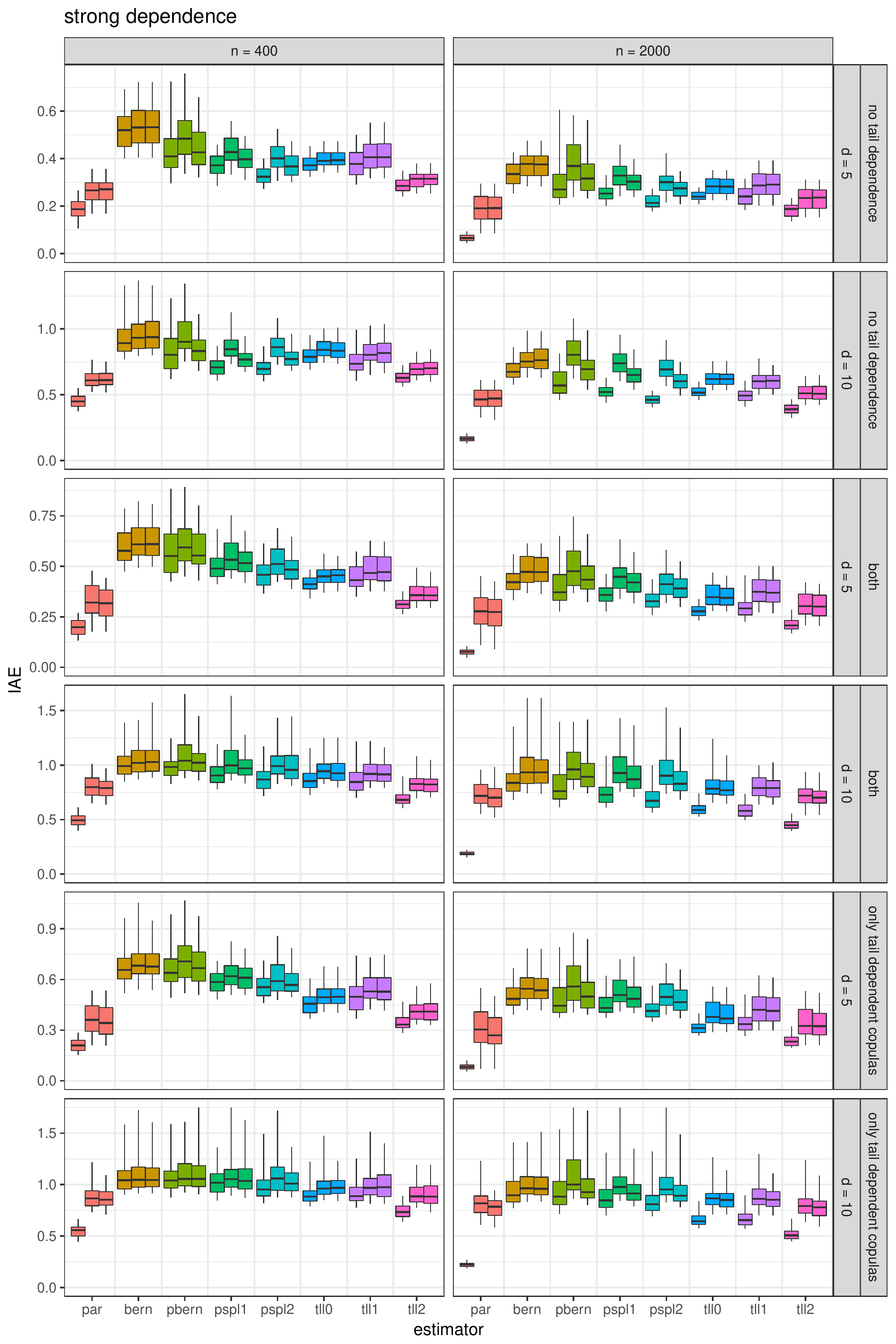} 
\caption{\emph{strong dependence:} the box plots show the IAE achieved by each estimation method. Results are split by sample size, dimension, and type of dependence.  Per estimator there are three boxes, corresponding to estimation under known structure, selection by Kendall's $\tau$, and selection by cAIC (from left to right).}\label{fig:strong}
\end{figure}

\autoref{fig:weak} and  \autoref{fig:strong} present the results of the simulation study described in \autoref{sec:simulations}. The analysis will be divided into several sections. The first takes a very broad view, whereas the remaining ones investigate the influence of individual factors. We acknowledge that the information density in these figures is extremely high. So we start with a detailed description of the figures' layout. 

\autoref{fig:weak} contains the results for all scenarios with weak dependence; \autoref{fig:strong} with strong dependence. The left columns correspond to the smaller sample size ($n=400$) and the right columns to the larger sample size ($n = 2\, 000$). The figures are also partitioned row-wise with an alternating pattern of the dimensions $d =5$ and $d=10$. Two subsequent rows correspond to the same type of dependence (no tail dependence, both, only tail dependence). In total there are 32 panels, each representing one of the 32 possible combinations of the factors listed in \autoref{sec:simulations:factors_tab}.

Each panel contains 24 boxes in 8 groups. Each group corresponds to one estimation method for the pair-copulas. The three boxes in each group represent the three different methods for structure selection: known structure, maximum spanning trees with Kendall's $\tau$, maximum spanning trees with cAIC (from left to right). The the box spans the interquartile range, the median is indicated by a horizontal line, the whiskers represent the 10\% and 90\% percentiles.

\subsection{Overall ranking of methods for pair-copula estimation}
\label{subsec:ranking}
We begin our analysis with a broad view on the relative performance of the pair-copula estimators. We want to assess the performance of the estimation methods, averaged over all scenarios and structure selection strategies. But just taking the average IAE could be misleading. It is evident from \hyperref[fig:weak]{Figures \ref*{fig:weak}} and \ref{fig:strong} that the scale of the IAE varies between scenarios. Averaging the bare IAE leads to an unbalanced few, laying more weight on particular scenarios. As a more robust alternative, we take the following approach: in each scenario, average the IAE over replications and structure selection strategies. Then rank the estimation methods by their relative performance. Ranks are comparable across scenarios, so our final criterion will be the average rank across all scenarios. These numbers are listed in \autoref{tab:av}.

\begin{table}[t]
\centering
\begin{tabular}{l|rrrrrrrr}
 & \texttt{par} & \texttt{bern} & \texttt{pbern} & \texttt{pspl1} & \texttt{pspl2} & \texttt{tll0} & \texttt{tll1} & \texttt{tll2} \\ 
  \hline
average rank & 1.28 & 7.17 & 6.35 & 5.00 & 5.01 & 5.01 & 3.83 & 2.35 \\
\hline
\end{tabular}
\caption{The relative rank of estimators averaged over all scenarios.}
\label{tab:av}
\end{table}

The parametric estimator performs best, which is no surprise since our simulation models consist of only parametric copula families. We included it in this study mainly to get a sense of what is possible in each scenario. Remarkably, it is outperformed in very few cases by a nonparametric estimator. This is due to the need for structure selection which will be discussed in more detail later on.\par
Among the nonparametric estimators, the kernel estimators (\texttt{tll2}, \texttt{tll1}, \texttt{tll0}) perform best, followed by the spline methods (\texttt{pspl1}, \texttt{pspl2}) which perform as well as the worst  kernel estimator \texttt{tll0}. The Bernstein estimators (\texttt{pbern}, \texttt{bern}) perform worst. Within these three classes, the accuracy improves mostly by how complex the estimation method is: going from regular Bernstein copulas to penalized ones; and going from local constant, to local linear, to local quadratic likelihood. It is the other way around for the B-spline methods, but the difference in the average rank is minuscule.

We will find that this relative ranking is fairly robust across scenarios. In the following analysis, we treat it as the benchmark ranking and focus on deviations from it.

\subsection{Strength and type of dependence}

By looking at the scale in each panel, we see that the performance of all estimators gets worse for increasing strength of dependence and increasing proportion of tail dependent families. This is explained by the behavior of the true densities. Many copula densities (and their derivatives) explode at a corner of the unit square. From the pair-copula families in our simulation model, only the Frank copula is bounded. Within each family, the tails explode faster when the strength of dependence increases. And tail dependence means that the tails explode particularly fast. Exploding curves are difficult to estimate for nonparametric estimators because their asymptotic bias and variance are usually proportional to the true densities' derivatives. Our results give evidence that this effect transfers to finite samples.

The estimators' response to these difficulties is the main driver behind their relative performance. In most scenarios, the ranking of estimators is similar to the benchmark rankings. But there are deviations. Let us walk through the scenarios one by one.

\begin{itemize}
\item \emph{weak, no tail dependence}: \texttt{pbern1} and \texttt{pspl1} perform better than \texttt{pspl2}, the kernel estimators, and even the parametric estimator for $n = 400$. For $n = 2\,000$, the parametric estimator gets ahead and the penalized methods are on par with \texttt{tll1} and \texttt{tll20}.

\item \emph{weak, both}:  \texttt{pbern1} and \texttt{pspl1} perform  better than \texttt{pspl2} and \texttt{tll0} for $n = 400$, and comparable for $n = 2\, 000$. 

\item \emph{weak, only tail dependent copulas}: similar to the benchmark ranking.

\item \emph{strong, no tail dependence}: \texttt{bspl2} beats \texttt{tll0} and \texttt{tll1} for $n=400$ and is on par for $n = 2\, 000$.

\item \emph{strong, both}: similar to benchmark ranking.

\item \emph{strong, only tail dependent copulas}: similar to benchmark ranking.
\end{itemize}

Overall, the penalized estimators tend to do better under weak dependence and only little tail dependence, whereas the kernel estimators do better in the other scenarios. The method \texttt{tll2} is the top performer in all but a few cases.

\subsection{Sample size and dimension}

When the sample size increases, the estimators become more accurate. Any reasonable estimator should satisfy this property. The kernel estimators and the non-penalized Bernstein estimator seem to benefit more from an increased sample size. The effect is most obvious in the weak dependence, no tail dependence case. This has an explanation: theoretically, the number of knots used by the penalized estimators should increase with the sample size. But our implementation uses a fixed number of knots, as the computational burden is already substantial compared with the other methods (see \autoref{subsec:time}). All other methods adapt their smoothing parameterization to the sample size. It is very likely that the penalized methods improve when the number of knots is further increased.

Comparing a pair of panels corresponding to $d=5$ and $d=10$, we see very little differences. We conclude that the results are very robust to changes in the dimensionality. 

\subsection{Structure selection}

The first aspect we want to discuss is the loss in accuracy caused by the need to select the tree structure. Recall that the three subsequent boxes for each estimator correspond to: estimation under the true structure (in practice unknown), selection based on Kendall's $\tau$, selection based on cAIC. 

The IAEs for the two selection methods are always higher than the `oracle' results with known structure. This makes sense: the true model is a simplified vine copula; if the true structure is known, the models are correctly specified and all estimators are consistent. In practice, the true structure is unknown, and a different structure will be selected most of the time. For the selected structure, there is no guarantee that the model is still simplified or that the estimators are consistent. For more details, we refer to \citet{Spanhel2015}, and \citet[Section 8]{Nagler15}.

Overall, the average loss in accuracy when going from the true to a heuristically selected structure increases with strength of dependence and prevalence of tail dependence. But the extent of this effect varies between estimation methods.
The parametric estimator suffers the most substantial losses. In fact, the the parametric estimator's performance is often very close to that of the best nonparametric estimator when the structure is unknown. This is quite remarkable considering that we simulate from parametric models. Interestingly, the loss for the penalized Bernstein and B-spline methods (\texttt{pbern}, \texttt{pspl1}, \texttt{pspl2}) is negligible in most scenarios when cAIC is used---but not when Kendall's $\tau$ is used. This is a distinct property of these penalized methods. The non-penalized Bernstein and kernel methods perform similarly for the two structure selection criteria. In most scenarios, the relative performance ordering of the estimators is the same for each type of structure. But there are a few cases (strong dependence, n = 400) where the \texttt{bspl2} estimator is worse than \texttt{tll0} or \texttt{tll1} with Kendall's $\tau$, but better with cAIC.

The results give evidence that the cAIC is the better criterion in terms of the estimators' accuracy. But  it also makes the vine copula estimators more costly to fit (see \autoref{subsec:structure}). So there is a trade-off between speed and accuracy. It usually depends on the problem at hand which to prioritize. We will investigate this issue further in the next section.
\subsection{Computation time}
\label{subsec:time}
\autoref{tab:time} lists the average computation time\footnote{The time was recorded on a single thread of a 8-way Opteron (Dual-Core, 2.6 GHz) CPU with 64GB RAM.} required to fit a vine copula and evaluate its density on 1\,000 importance Monte-Carlo samples. The results are divided into the combinations of dimension $d$ and sample size $n$.

\begin{table}[t]
\centering
\begin{tabular}{rrl|rrrrrrrrr}
  \hline
d & n & criterion & par & bern & pbern & pspl1 & pspl2 & tll0 & tll1 & tll2 \\ 
  \hline
5   & 400 & $\tau$ &  7 & 3 & 788 & 758 & 517 & 3 & 4 & 6 \\ 
   &    & cAIC & 19 & 10 & 1\,000 & 1\,175 & 786 & 10 & 11 & 13 \\ 
  &2\,000 & $\tau$ &  34 & 19 & 1\,578 & 1\,455 & 1\,394 & 7 & 12 & 16 \\ 
 & & cAIC & 91 & 31 & 2\,163 & 2\,336 & 2\,243 & 25 & 32 & 35 \\ 
10 & 400 & $\tau$ & 33 & 17 & 2\,983 & 3\,183 & 2\,205 & 14 & 19 & 29 \\ 
   & & cAIC & 98 & 49 & 5\,292 & 6\,110 & 4\,156 & 48 & 55 & 65 \\ 
 & 2\,000 & $\tau$ & 159 & 65 & 6\,553 & 6\,694 & 6\,515 & 35 & 56 & 71 \\ 
 & & cAIC & 472 & 139 & 11\,992 & 13\,514 & 12\,394 & 127 & 158 & 173 \\ 
   \hline
\end{tabular}
\caption{Average computation time (in seconds) required for estimation and selection of one vine copula model.}
\label{tab:time}
\end{table}
Let us first focus on the selection criterion. We clearly see that the computation time increases substantially for all estimators when cAIC is used instead of Kendall's $\tau$. This effect size differs, but is usually a factor of around two or three.

The fastest two estimators are the simplest ones: \texttt{bern} and \texttt{tll0}. The other two kernel estimators are in the same ballpark, but the computation time increases slightly with the order of the polynomial. Only slightly slower is the parametric estimator. The reason is that the parametric estimator has to iterate through several different copula families before it can select the final model. The penalized estimators are about orders of magnitude slower than their competitors. Take for example the case $d = 10$ and $n = 2\,000$, where most estimators take around one minute (using $\tau$), but the penalized estimators take more than one and a half hours.  

The large difference in computational demand is caused by the penalized estimation problem. One has to optimize over more than 100 parameters with more than 100 side constraints. Even worse, such a problem has to be solved multiple times until an optimal choice for the penalty parameter $\lambda$ has been found. Reducing $K$ (the number of knots) does significantly reduce this burden, but also limits the flexibility of the estimators. In the end, the statistician has to choose which $K$ yields the best balance between speed and accuracy.

\subsection{Limitations}

The referees pointed out some limitations of our study which are addressed in the following.

\subsubsection{Performance measure}

All results focus on a single performance measure and therefore only provide a limited view on the estimators' performance. Although this is true, we considered several other measures in preliminary versions of this study and found the results to be quite robust w.r.t.\ to the measure.

\subsubsection{Estimation of marginal distributions}

The study neglects the fact that observations from the copula are never observed and one has to rely on pseudo-observations that depend on estimated marginal distributions. \citet{Kim2007} showed in an extensive simulation study that this can be a problem in misspecified fully parametric models. But the issue is largely resolved when the margins are estimated nonparametrically. In this case, maximum likelihood estimators are unbiased and only slightly less efficient \citep[][]{genest1995}. 

In a purely nonparametric context, it is even less of an issue. In fact, many authors have found that errors stemming from estimating the marginal distributions are asymptotically negligible when estimating the copula density \citep[e.g.,][]{Janssen2014, Geenens14a, Nagler16}. This is explained by the fact that distribution functions can be estimated consistently at the parametric rate, whereas density estimators are bound to (slower) nonparametric rates. Accordingly, we can expect similar results to the ones presented even if margins were treated unknown.

\subsubsection{Choice of smoothing parameters}

It is a common theme in nonparametric estimation that the quality of estimators depends heavily on the choice of smoothing parameters. This is certainly also the case for the estimators considered in this study. However, we do not think it is feasible to assess the sensitivity of our results to this choice: 
\begin{itemize}
  \item Smoothing parameters are hardly comparable across estimation methods because they arrive at the density estimate in fundamentally different ways.

  \item There are too many smoothing parameters in a vine copula model: There are 10 ($d = 5$) resp.\ 45 ($d = 10$) pair-copulas, and for each pair-copula there are between one and three smoothing parameters (depending on the estimation method).

  \item Due to the sequential nature of the joint estimator, pair-copula estimators in later trees are affected by the estimates in earlier trees. This leads to significant interactions between smoothing parameters at different levels.
\end{itemize}
In our study, all smoothing parameters were selected by automatic procedures that are state-of-the-art. This realistically reflects statistical practice. But one should keep in mind that the performance of most estimators can likely be improved by advances in automatic selection procedure.

    \section{Illustration with real data}
\label{sec:illustration}

In the simulation study, the parametric estimator performed best in virtually all scenarios. But this is simply a consequence of simulating from parametric models. But real data does not always behave that nicely and nonparametric methods are required to appropriately capture the dependence. Such a case is illustrated in the following real-data example.

We consider a data set representative of measurements taken on images from the MAGIC (Major Atmospheric Gamma-ray Imaging Cherenkov) Telescopes\footnote{https://archive.ics.uci.edu/ml/datasets/MAGIC+Gamma+Telescope} with 19020 observations, focusing on gamma observations. It has been previously analyzed with a kernel-based simplified vine copula estimator by \citet{Nagler15}. To show exemplary results of the different nonparametric copula density estimators, we select a random subset ($n=2000$) from the MAGIC data with respect to to the variables fConc1, fM3Long and fM3Trans. We compute pseudo-observations from the data by applying the marginal empirical distribution functions to each variable.

\begin{figure}[!t]
\begin{center}
\includegraphics[width =\textwidth]{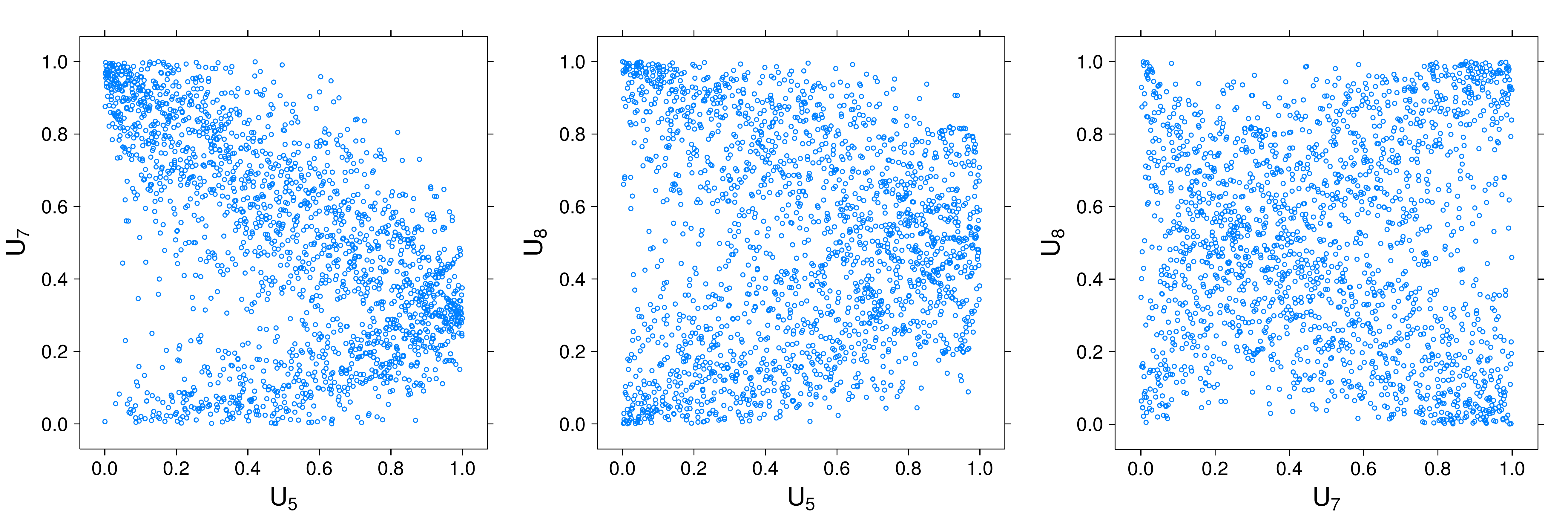}
\end{center}
\caption{Scatterplots of pseudo observation ranks for pairs (left) fConc1 ($U_5$) and fM3Long ($U_7$), (middle) fConc1 ($U_5$) and fM3Trans ($U_{8}$) and (right) fM3Long ($U_7$) and fM3Trans ($U_{8}$) from the MAGIC data set ($n=2000$).} \label{pic:pairs}
\end{figure}

\begin{figure}
\begin{center}
\vspace{-0.25cm}
\includegraphics[width=\textwidth]{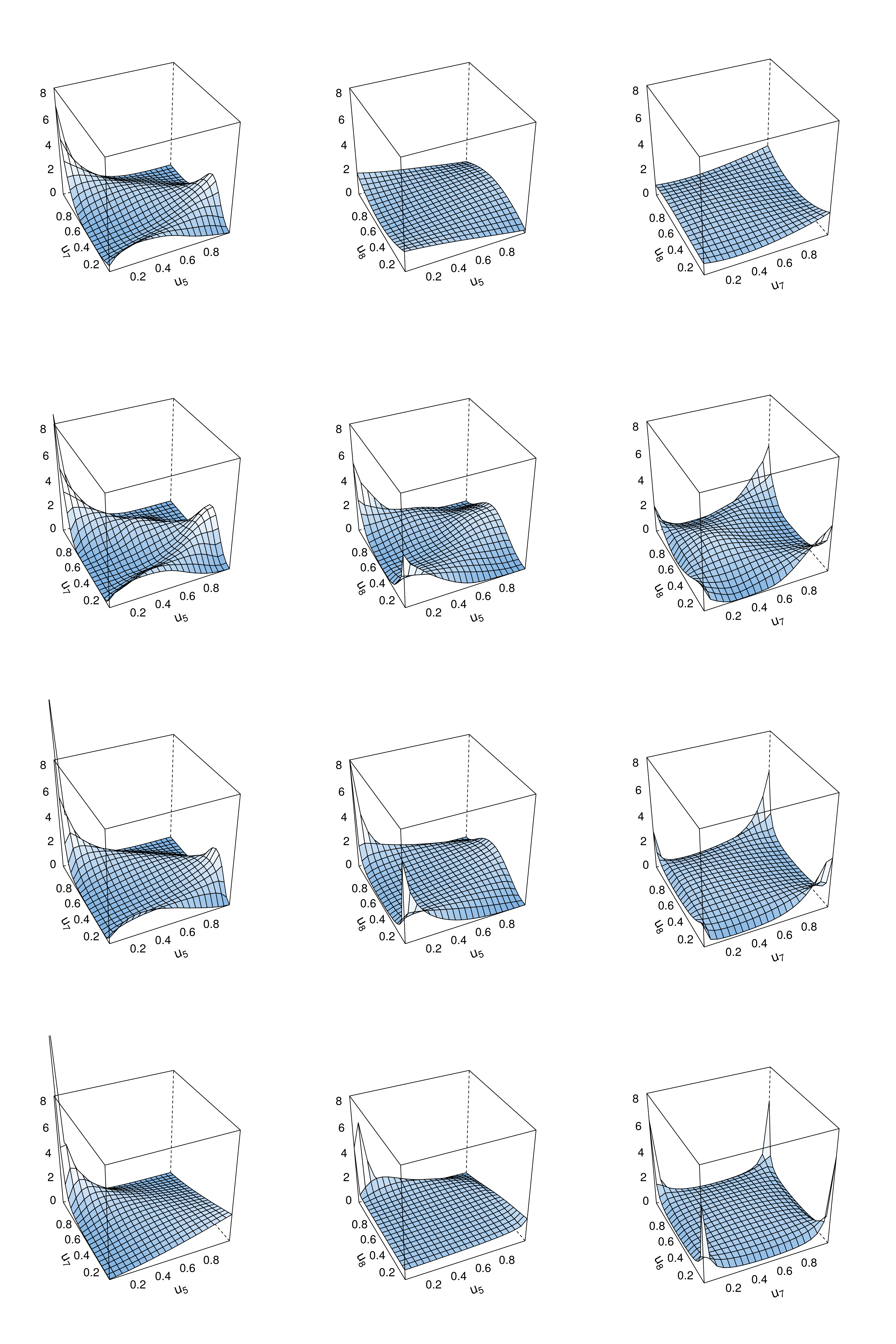}
\end{center}
\caption{Exemplary density plots for MAGIC data ($n = 2000$). 1st row: Bernstein estimator \texttt{bern}, 2nd row: penalized quadratic B-splines estimator \texttt{pspl2}, 3rd row: kernel estimator \texttt{tll2}, 4th row: parametric estimator \texttt{par}.}\label{pic:magic.ex}
\end{figure}

\autoref{pic:pairs} shows scatter plots of the three pairs of the pseudo-observations. The shapes we see are different from what we know from popular parametric families.
We fit several copula density estimators to each pair and show the results in \autoref{pic:magic.ex}. The first column of \autoref{pic:magic.ex} shows the fitted pair-copula density between fConc1 ($U_5$) and fM3Long ($U_7$), the second column between fConc1 ($U_5$) and fM3Trans ($U_{8}$) and the third column contains the copula density between fM3Long ($U_7$) and fM3Trans ($U_{8}$). \par

The first pair of variables fConc1 $(U_5)$ and fM3Long $(U_7)$ a lot of pseudo-observations accumulate around the point $(0,1)$, which is reflected as high density peaks in all fitted copula densities. But for the accumulation around the point $(1, 0.3)$, we observe a difference between the nonparametric estimators \texttt{bern}, \texttt{pspl2} and \texttt{ttl2} and the parametric copula density, which does not mirror this accumulation.\par

For the second pair, fConc1 $(U_5)$ and fM3Trans $(U_8)$, the estimated density varies considerably between methods. Estimates of \texttt{pspl2} and \texttt{ttl2} show  peaks around the points $(0,0)$ and $(0,1)$, which reflects the the large concentration of points in the scatter plot in \autoref{pic:pairs}. The estimators \texttt{bern} and \texttt{par} do not contain these peaks. We observe similar differences for the estimated densities for the third data pair, presented in the right column of \autoref{pic:pairs}. While \texttt{bern}, \texttt{pspl2} and \texttt{ttl2} show density peaks around the accumulation points $(1,0)$ and $(1,1)$, but the estimated parametric copula does not map this structures of the data.

The previous examples have illustrated situations parametric estimator fails because of its lack of flexibility. In such situations, nonparametric methods are required to adequately capture the true dependence structure. However, we merely looked at two unconditional pairs of variables, not a full dependence model.

\begin{figure}
\begin{center}
\includegraphics[width = \textwidth]{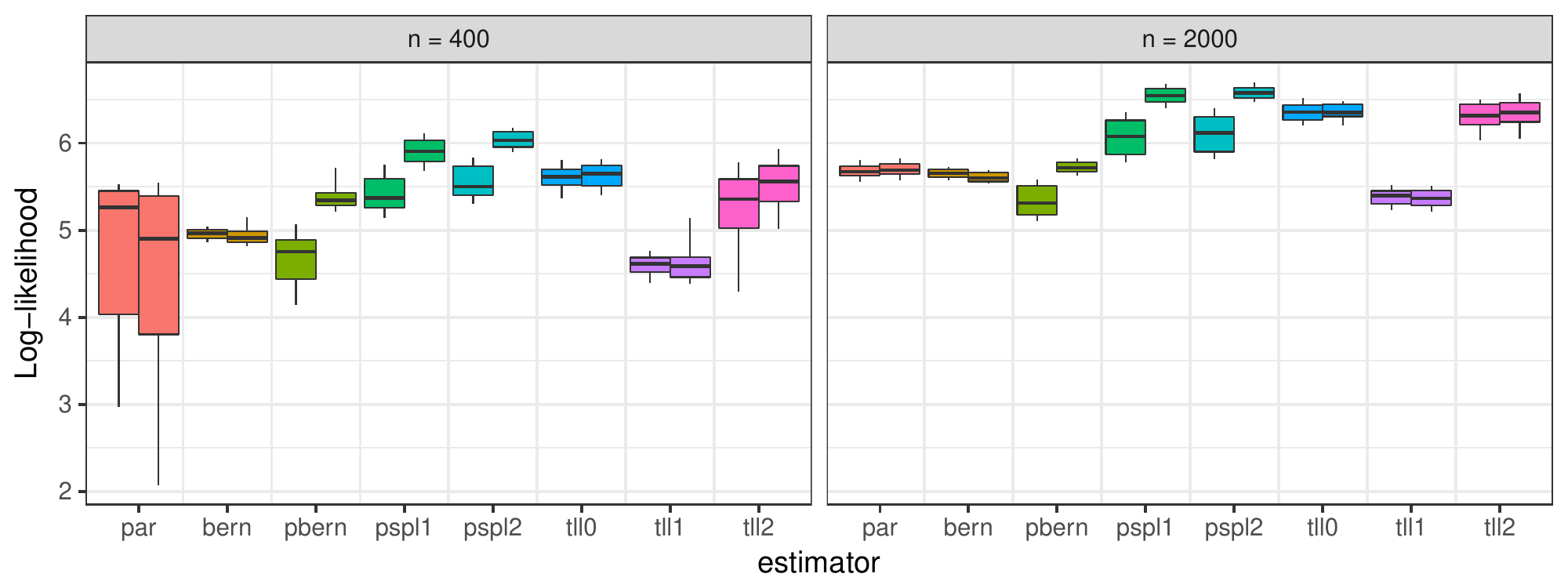}
\end{center}
\caption{The box plots show the mean log-likelihood values attained by the different estimation methods. Each boxplot on the left hand side: structure selection based on Kendell's $\tau$ and each boxplot on the right hand side: structure selection based on cAIC.}\label{pic:magic}
\end{figure}
Similar to our simulation study, we assess the performance of the estimators via cross-validation. We randomly draw a size $n$ subset $\bm{U}_{\mathrm{train}}$ of the data, apply the estimators and calculate the mean out-of-sample log-likelihood on $1\,000$ randomly selected remaining observations $\bm{{U}}_{\mathrm{test}}$, i.e.,
\begin{equation*}
  \ell(\bm{{U}}_{\mathrm{test}})=\frac{1}{1000}\sum_{i=1}^{1000} \ln \hat{c}(\bm{{U}}_{\mathrm{test}}^{(i)}),
\end{equation*}
where $\hat{c}$ is a vine copula density estimator based on $\bm{U}_{\mathrm{train}}$. This is repeated $N=100$ times for sample sizes $n=400$ and $n=2\,000$.
The results are summarized as box plots in \autoref{pic:magic} for all estimators and structure selection based on Kendall's $\tau$ (left box) and cAIC (right box).\par
The parametric estimator performs unsatisfactory for $n=400$ since it varies enormously for both data sets. But also for $n=2000$, the parametric estimator is outperformed by most nonparametric alternatives. The performance of the nonparametric methods varies notably between methods. The methods \texttt{bern} and \texttt{tll1} do not perform well, but the other methods clearly outperform \textrm{par}. Furthermore, the performance differs significantly with respect to the structure selection criterion for \texttt{bern1} and \texttt{pspl1, pspl2}. For small sample size ($n=400$) and using Kendall's $\tau$ as selection criterion, \texttt{tll0} results with highest mean, directly followed by \texttt{pspl2}. But choosing cAIC instead, the log-likelihood of \texttt{pspl2} increases, but not for \texttt{tll0}. The situation is similar for $n=2000$.
We conclude that the more sophisticated nonparametric methods adequately reflect the distribution of the data. In contrast, the dependence structure observed in \autoref{pic:pairs} can not be captured adequately with standard parametric models.
    \section{Conclusion}
\label{sec:conclusion}

This articled compared existing methods for nonparametric estimation of simplified vine copula densities. The estimators considered are the non-penalized Bernstein estimator, the penalized Bernstein estimator, penalized B-spline estimators (linear and quadratic), and kernel weighted local likelihood estimators (local constant, linear, and quadratic). We compared these methods by an extensive simulation study and on two real data sets.

The simulation study comprises several scenarios for sample size, dimension, strength of dependence, and tail dependence. The simulation models are set up as parametric vine copulas with randomized vine structure, pair-copula families, and parameters. Overall, the kernel methods were found to perform best (especially the local quadratic version), followed by the penalized B-spline estimators. The Bernstein estimators performed worst. An exception to this pattern was found in scenarios with small sample size, weak dependence, and no tail dependence. Here, the penalized B-spline and Bernstein estimators outperformed the kernel methods. Additionally, we demonstrated the need for nonparametric methods on real data whose dependence structure that cannot be adequately captured by a the parametric estimator.

Overall, we found that no estimator is uniformly better than the others; it depends on the data which is to be preferred. Our analysis highlighted which factors drive the performance of the various methods, and which methods should be preferred for certain scenarios. In applications, statisticians can determine the characteristics of their data by an exploratory analysis, and make a well-informed choice based on these results.

\subsection*{Acknowledgements}
The first and third author were partially supported by the German Research Foundation (DFG grants CZ  86/5-1 and CZ 86/4-1). All authors thank the Leibniz Supercomputing Centre  (\url{www.lrz.de}) for providing access to a computing cluster.
    %\input{appendix}

	% load references
	\bibliography{References}
	
\end{document}